\documentclass[prl,twocolumn,showpacs,superscriptaddress]{revtex4}

\usepackage{graphicx,color}
\usepackage{textcomp}
\usepackage{amsmath}

\newcommand{\unit}[1]{\ensuremath{\,\mathrm{#1}}} 
\newcommand{\degc}{\ensuremath{\,^\circ\mbox{C}}}

\begin{document}

\title{Coherence measures for heralded single-photon sources}

\author{E. Bocquillon}
\affiliation{Institute for Quantum Computing and Department of Physics and Astronomy,
\\University of Waterloo, 200 University Ave W., Waterloo, ON, Canada  N2L 3G1}
\affiliation{Ecole Normale Sup\'{e}rieure, 45, rue d'Ulm, 75230 Paris, France}
\author{C. Couteau}
\affiliation{Institute for Quantum Computing and Department of Physics and Astronomy, \\University of Waterloo, 200 University Ave W., Waterloo, ON, Canada  N2L 3G1}
\author{M. Razavi}
\email{mrazavi@iqc.ca}
\affiliation{Institute for Quantum Computing and Department of Electrical and Computer Engineering, \\University of Waterloo, 200 University Ave W., Waterloo, ON,
Canada  N2L 3G1}
\author{R. Laflamme}
\affiliation{Institute for Quantum Computing and Department of Physics and Astronomy,
\\University of Waterloo, 200 University Ave W., Waterloo, ON, Canada  N2L 3G1}
\affiliation{Perimeter Institute for Theoretical Physics,
31 Caroline Street N., Waterloo, ON, Canada N2L 2Y5}
\author{G. Weihs}
\affiliation{Institute for Quantum Computing and Department of Physics and Astronomy,
\\University of Waterloo, 200 University Ave W., Waterloo, ON, Canada  N2L 3G1}

\begin{abstract}
Single-photon sources (SPSs) are mainly characterized by the minimum value of their second-order coherence function, viz. their $g^{(2)}$ function. A precise measurement of $g^{(2)}$ may, however, require high time-resolution devices, in whose absence, only time-averaged measurements are accessible. These time-averaged measures, standing alone, do not carry sufficient information for proper characterization of SPSs. Here, we develop a theory, corroborated by an experiment, that allows us to scrutinize the coherence properties of heralded SPSs that rely on continuous-wave parametric down-conversion. Our proposed measures and analysis enable proper standardization of such SPSs.
\end{abstract}
\maketitle

The demand for ultra-secure communication, high-precision measurement, and super-efficient computation \cite{Bennett84a} has resulted in the emergence of optical sources that create single---and, ideally, only single---photons in a heralding and/or on demand way \cite{HSPS}. The rapid progress in this area has been followed by its early introduction to the commercial market \cite{market}, even before finalizing a proper set of standards for characterizing such devices. For a heralded single-photon source (HSPS) that relies on the spontaneous parametric down-conversion (SPDC), where the detection of idler photons heralds for the presence of signal photons, two figures of merit are generally of crucial importance. The first is the temporal correlation between the signal and idler beams \cite{valencia01a}, on which our heralding mechanism relies, and the second is the second-order degree of coherence for the heralded signal photons \cite{rarity87a}. The challenge of measuring either of these figures lies in the large bandwidth of the SPDC process, thereby the very narrow widths of such correlation functions. In fact, what we can commonly measure in an experimental setup is only a time-averaged version of the actual figure. It is important then to recognize all major parameters that affect our measurement results, and put them together to come up with well-defined, and readily measurable, figures of merit for HSPSs. This Report carves into the theoretical aspects of such problems and addresses the above coherence measures, and their corresponding time-averaged figures, with an accuracy never presented before. Our analysis accounts for the contribution of multi-photon states in the SPDC process as a function of pump power, or, effectively, the single-photon generation rate, for different widths of the coincidence window and photodetectors' time resolutions. Our theoretical predictions are corroborated by our experimental results, and they together provide a prescription for proper characterization of HSPSs.

\begin{figure}
\begin{center}
\includegraphics [width=.95\linewidth] {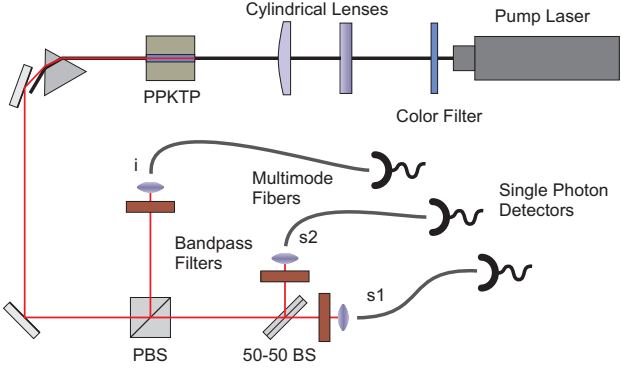}
\caption{(Color online) Experimental setup for our heralded single-photon source. A blue laser is cleaned from infrared fluorescence by a color filter, is focused by cylindrical lenses onto a PPKTP crystal (cut and poled for type-II), and is then removed from the down-converted beam by a dispersive prism. A polarizing beam splitter (PBS) splits the down-converted beam into the signal and idler arms. The idler beam i is used as a trigger and, in an HBT interferometer, the signal beam is split by a 50:50 beam splitter into s1 and s2 for the coherence function measurement. Bandpass filters block background light from entering the multi-mode fibers that couple the light to the single-photon detectors.
\label{Fig:setup}}
\end{center}
\end{figure}

Figure~\ref{Fig:setup} shows the setup of our HSPS along with the Hanbury Brown and Twiss (HBT) interferometer used for the measurement of the coherence function. In our experiment, a 405\unit{nm} continuous-wave laser 
pumped a periodically poled $\mathrm{KTiOPO_4}$ (PPKTP) crystal. The crystal from Raicol was cut to $10\times2\times1\unit{mm^3}$ for propagation along the $x$ axis and poled with a $10\,$\hbox{\textmu}m period to support type-II SPDC, where signal and idler photons have orthogonal polarizations. By using an oven, frequency degeneracy was reached at $39\degc$ with a stability of $\pm 0.1\degc$. A polarizing beam splitter (PBS) split the two beams into two different spatial modes. A photodetection event on the idler beam heralded the presence of one or more photons on the signal beam, which went through an HBT interferometer consisting of a 50:50 beam splitter followed by two bandpass filters (810\unit{nm}, 10\unit{nm} bandwidth) and multi-mode fibers that coupled the light to photodetectors. We employed single-photon counting modules from Perkin-Elmer with nominal dead-times of 45\unit{ns}. Typical measured photon count rate for our setup was up to 800,000 counts/s in the idler channel, with a signal-idler coincidence rate amounting up to 10\% of that value.

The detection times for the signal and idler beams were recorded by a time-tagging unit from Dotfast Consulting with a nominal temporal resolution of 156.25\unit{ps}. The time-tagging unit streamed the time tags to a computer, by which we could record single, double, and triple detection events between the three channels, $i$, $s_1$, and $s_2$, in Fig.~\ref{Fig:setup}. Coincidence windows were implemented only in software. The complete system of photodetectors, power supplies, time-stamping electronics, and the USB interface fit in a $30\times30\times30\unit{cm^3}$ box.

The first coherence measure studied here is the temporal correlation between signal and idler beams, as a measure of reliability of our heralding mechanism, and is defined as follows
\begin{equation}
\label{Eq:g_si_2}
g_{si}^{(2)}(t,\tau)  \equiv  \frac {\langle \hat E_s^\dag(t+\tau) \hat E_i^\dag(t) \hat E_i(t) \hat E_s(t+\tau) \rangle} {\langle \hat E_s^\dag(t+\tau) \hat E_s(t+\tau) \rangle  \langle \hat E_i^\dag(t) \hat E_i(t) \rangle},
\end{equation}
where $\hat E_s(t)$ and $\hat E_i(t)$ represent the scalar photon-units positive-frequency field operators for the outgoing signal ($s$) and idler ($i$) beams, respectively. The joint state of signal and idler is a zero-mean Gaussian state whose only nonzero second-order moments are given by \cite{Wong06a}
\begin{eqnarray}
\label{Eq:autocor}
 & e^{i \omega_p \tau / 2} R(\tau) \equiv \langle \hat E_k^\dag(t+\tau) \hat E_k(t) \rangle , \quad\mbox{$k=s,i$},&\\
\label{Eq:crosscor}
& e^{-i \omega_p (t+\tau)/2}  C(\tau) \equiv \langle \hat E_s(t+\tau) \hat E_i(t) \rangle ,&
\end{eqnarray}
which represent the auto-correlation function and the
phase-sensitive cross-correlation function between signal and idler
fields, respectively, and where $\omega_p$ is the pump frequency. All
other moments can be obtained by using the quantum form of the
Gaussian moment-factoring theorem \cite{Shapiro94a}. For instance,
the numerator in Eq.~(\ref{Eq:g_si_2}) can be simplified as follows
\begin{eqnarray}
\label{Psit1t2}
P_{si}(\tau) & \equiv & {\langle \hat E_s^\dag(t+\tau) \hat E_i^\dag(t) \hat E_i(t) \hat E_s(t+\tau) \rangle} \nonumber \\
& = & {\langle \hat E_s^\dag(t+\tau) \hat E_i^\dag(t) \rangle \langle \hat E_i(t) \hat  E_s(t+\tau) \rangle} \nonumber \\
& + &  {\langle \hat E_s^\dag(t+\tau) \hat E_i(t) \rangle \langle \hat E_i^\dag(t)  \hat  E_s(t+\tau) \rangle} \nonumber \\
& + & {\langle \hat E_s^\dag(t+\tau)  \hat  E_s(t+\tau) \rangle \langle \hat E_i^\dag(t) \hat E_i(t)  \rangle} \nonumber \\
& = & R^2(0) + |C(\tau)|^2 ,
\end{eqnarray}
which represents the coincidence rate for having a signal photon at time $t+\tau$ and an idler photon at time $t$.

In the low-gain regime of parametric down-conversion, which prevails in our case, the auto- and cross-correlation functions can be approximated by the following expressions \cite{Wong06a}
\begin{equation}
\label{Rcorr}
R(\tau) = \left\{\begin{array}{cc}
R_{\rm SPDC} (1+ \tau / \Delta {\rm t} ) & -\Delta {\rm t} < \tau \leq 0 \\
R_{\rm SPDC} (1- \tau / \Delta {\rm t}) & 0 < \tau \leq \Delta {\rm t} \\
0 & \mbox{elsewhere}
\end{array} \right. ,
\end{equation}
where $R_{\rm SPDC}$ is the rate of photon generation for the signal/idler
beam and $1/\Delta {\rm t}$ is the bandwidth of the SPDC process,
and
\begin{equation}
\label{Ccorr}
|C(\tau)| = \left\{\begin{array}{cc}
\sqrt{R_{\rm SPDC} / \Delta {\rm t}}  & -\frac{\Delta {\rm t}}{2} < \tau < \frac{\Delta {\rm t}}{2} \\
0 & \mbox{elsewhere}
\end{array} \right. ,
\end{equation}
with the difference in the speed of light for ordinary and extraordinary axes in the crystal being compensated.

From Eqs.~(\ref{Eq:g_si_2}) and (\ref{Psit1t2}), we have $ g_{si}^{(2)}(t,\tau) = 1 + |C(\tau)/R(0)|^2$. For our source, at 50\unit{mW} pump power, $R(0) = R_{\rm{SPDC}} \approx 43$\unit{MHz} and $1/\Delta {\rm t} \approx 3$\unit{THz}, which results in a peak value of $1+1/(\Delta {\rm t} R_{\rm{SPDC}}) \approx 7\cdot10^4$ for $g_{si}^{(2)}(t,\tau)$ at $\tau =0$. The coherence function quickly drops to its minimum value one within a sub-picosecond period, however the finite time resolution in our experiment will smooth this feature out as we show next.

In order to measure $ g_{si}^{(2)}(t,\tau)$, we approximate $P_{si}(\tau)$ by the rate of coincident events, $N_{si} (\tau)$, in which an idler photocount is observed at time $t$ and a signal photocount is observed in the interval $[t+\tau-\tau_{\rm coin},t+\tau+\tau_{\rm coin}]$, where $2 \tau_{\rm coin}$ is the width of our chosen coincidence window. Because of the photodetectors' time jitters, and neglecting dark counts throughout the Report, a photodetection event at time $t$ only implies the existence of one or more photons in a neighborhood around time $t$. For simplicity, we assume that the detection time corresponding to a photon that hits the detector's surface at time $t$ is uniformly distributed over the interval $[t-\tau_d,t+\tau_d]$, where $\tau_d$ is the time resolution of the photodetectors. We can then write the observed value for $N_{si} (\tau)$ in terms of $P_{si}(\tau)$ in the following way \cite{note1}
\begin{equation}
\label{Nsi}
N_{si}(\tau)  \approx \frac{1}{2 \tau_{\rm coin}} \int_{\tau-\tau_{\rm coin}}^{\tau+\tau_{\rm coin}}{d \tau' \bar P_{si}(\tau')}  ,
\end{equation}
where
\begin{equation}
\label{barPsi}
\bar P_{si}(\tau) = \int{d t_i \int{d t_s u(t_i) u(t_s-\tau) P_{si}(t_s-t_i)}}
\end{equation}
is the coincidence rate for detecting a signal photon at time $t+\tau$ and an idler photon at time $t$, where $u(t) = 1/(2 \tau_d)$ if $|t|\leq \tau_d$, and zero otherwise.

\begin{figure}
    \centering
    \includegraphics[width = .95\linewidth] {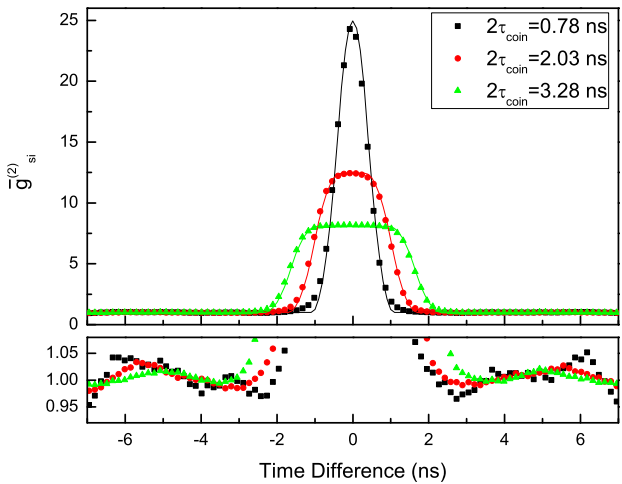}
    \caption{(Color online) \textbf{Top:} Measurements (symbols) and theory predictions (lines) of the time-averaged coherence function $\bar g_{si}^{(2)}(\tau)$ for the signal and idler photons. The low-gain regime theory curves are in striking agreement with the data using the following parameter values $R_{\rm{SPDC}}=43\unit{MHz}$, $\tau_d = 350\unit{ps}$ and $1/\Delta {\rm t}=3\unit{THz}$, where the last one was measured separately by spectroscopy and the other two were adapted to provide the subjective best visual fit simultaneously to all three sets of data.
    \textbf{Bottom: }Same data as top but with magnified ordinate in order to reveal structure that is caused by double reflection of photons from the fiber input endface and the surface of another optical element in the experiment. While this effect causes only very small deviations (0.2\% of the central peak) from the expected flat line, these deviations cause noticeable ringing of the measured $\bar g_c^{(2)}$ (see Fig.~\ref{Fig:g2dip}). 
    }
    \label{Fig:g2corr}
\end{figure}

Figure~\ref{Fig:g2corr} shows the experimental and the theoretical results for the time-averaged coherence function $\bar g_{si}^{(2)}(\tau) \equiv N_{si}(\tau)/R^2(0)$ for different values of $\tau_{\rm coin}$. Experimentally, $R^2(0)$ was determined by the product of the signal and idler count rates. For the theoretical graphs, we used the low-gain correlation functions given by Eqs.~(\ref{Rcorr}) and (\ref{Ccorr}) with $R_{\rm SPDC}=43$\unit{MHz} and $1/ \Delta {\rm t} = 3$\unit{THz}. From Eqs.~(\ref{Ccorr})--(\ref{barPsi}), we see that $N_{si}(\tau)$ has an almost fixed value for $\tau \in [-\tau_{\rm coin}+\Delta {\rm t} +\tau_d , \tau_{\rm coin}-\Delta {\rm t}-\tau_d]$, inversely proportional to $\tau_{\rm coin}$. As we get farther from the center, the time-averaged coherence function drops to its minimum value one as expected. The theoretical graphs are in striking agreement with our experimental results, which clearly demonstrate the strong temporal correlation between signal and idler beams.

The second coherence measure that we consider here is the degree of second-order coherence for the signal field, conditioned on observing an idler photocount at time $t_i$, defined as follows
\begin{equation}
\label{Eq:Cond_g2}
g_c^{(2)}(t_1,t_2|t_i) \equiv \frac{\langle \hat E_s^\dag(t_1) \hat E_s^\dag(t_2) \hat E_s(t_2) \hat E_s(t_1) \rangle_{\rm pm}}{\langle \hat E_s^\dag(t_1) \hat E_s(t_1) \rangle_{\rm pm}  \langle \hat E_s^\dag(t_2) \hat E_s(t_2) \rangle_{\rm pm}},
\end{equation}
where $\langle \cdot \rangle_{\rm pm}$ is the average over the post-measurement state assuming sufficiently high time resolution and unity quantum efficiency for the idler photodetector.

To model the measurement on the idler field operator, we use a heuristic approach in which a photodetection event at time $t_i$ on the idler beam is modeled by the measurement operator,  \cite{Nielsen00a}, $\hat E_i(t_i)$. We can show that if we allow for infinitely high time resolutions this method provides us with the correct result \cite{Razavi09a}. The post-measurement averaging, for any operator $\hat X$, will then be given by
\begin{equation}
\langle \hat X \rangle_{\rm pm} = {\langle \hat E_i^\dag(t_i) \hat X \hat E_i(t_i)\rangle} / { {\langle  \hat E_i^\dag(t_i)\hat E_i(t_i)\rangle}} .
\end{equation}
The conditional coherence function in Eq.~(\ref{Eq:Cond_g2}) can then be written as follows
\begin{equation}
\label{Eq:gct1t2}
g^{(2)}_c(t_1,t_2|t_i)=\frac{P_{si}^{(2)}(t_1,t_2,t_i) R(0)}{P_{si}(t_1-t_i)P_{si}(t_2-t_i)},
\end{equation}
where, using the quantum version of the Gaussian moment-factoring theorem along with Eqs.~(\ref{Eq:autocor}) and (\ref{Eq:crosscor}),
\begin{eqnarray}
\label{Psit1t2ti}
P_{si}^{(2)}(t_1,t_2,t_i) & \equiv & \langle \hat{E}_i^\dag(t_i)\hat{E}_{s}^\dag(t_1)\hat{E}_{s}^\dag(t_2)\hat{E}_{s}(t_2)\hat{E}_{s}(t_1)
\hat{E}_i(t_i) \rangle \nonumber \\
& = & R(0)\left[R^2(0)+|R({t_1 - t_2})|^2 \right] \nonumber \\
& + & 2\Re \left\{C(t_1 -t_i)C^\ast(t_2 -t_i)R(t_1 -t_2)\right\} \nonumber \\
& + & R(0)\left[|C(t_1 -t_i)|^2 + |C(t_2 -t_i)|^2 \right]
\end{eqnarray}
is the multi-coincidence rate for having signal photons at times $t_1$ and $t_2$ and an idler photon at $t_i$.

The first of several interesting special cases we consider is the coherence function at the trigger time, i.e.,
\begin{equation}
\label{gc20}
g^{(2)}_c(t_i,t_i|t_i)  =  2 - \frac{2 |C(0)|^4  }
{\left(R^2(0)+ |C(0)|^2\right)^2}.
\end{equation}
It is clear that if $R^2(0) \ll |C(0)|^2$ then
$g^{(2)}_c(t_i,t_i|t_i) \approx 0$. This is the same requirement
that we had for observing a large $g_{si}^{(2)}(t,0)$ in
Eq.~(\ref{Eq:g_si_2}), and, therefore, a low value for
$g^{(2)}_c(t_i,t_i|t_i)$ is guaranteed if $g_{si}^{(2)}(t,0) \gg 1$.
The second interesting case is when $t_1 = t_i$ but $ |t_2-t_i| \gg 2 \Delta t$. In this case,
$g^{(2)}_c(t_i,t_2|t_i) \approx 1$ provided that $R^2(t_2-t_i) \approx 0$ and $|C^2(t_2-t_i)| \approx 0$. This implies that our HSPS has a coherence time on the order of $\Delta t$.
Finally, let us consider the case when $|t_1-t_i = t_2-t_i| \gg 2 \Delta t$, i.e, when there is no correlation between the trigger time and the signal beam. In this case,
$g^{(2)}_c(t_1,t_2|t_i) \approx 2$, which is expected because, in the lack of any triggering event, both signal and idler beams individually obey the thermal-state statistics, for which the second-order coherence function has a maximum value of two \cite{Loudon83a}.

\begin{figure}
\begin{center}
\includegraphics [width=.89\linewidth]{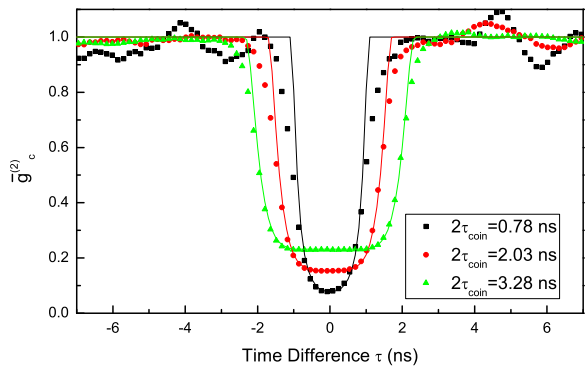}\caption{(Color online) Measured (symbols) and calculated (lines) time-averaged conditional coherence function $\bar g_c^{(2)}(\tau)$. The theory lines were calculated using the same parameter values as in Fig.~\ref{Fig:g2corr}. The purely statistical errors of our data are on the order of the symbol size in the figure and therefore not shown. As explained in the caption of Fig.~\ref{Fig:g2corr} photons that are reflected twice cause the apparent ringing.}
\label{Fig:g2dip}
\end{center}
\end{figure}

To quantitatively characterize our HSPS, it is interesting to measure $g_c^{(2)}(\tau) \equiv g_c^{(2)}(t_i, t_i+\tau|t_i) = g_c^{(2)}(0, \tau|0)$. For an ideal HSPS, we expect that $g_c^{(2)}(0)=0$. In our case, from Eq.~(\ref{gc20}), $g_c^{(2)}(0)\approx 6\cdot10^{-5} \ll 1$. Here, we measure a time-averaged version of the coherence function by approximating $P_{si}(\tau)$ with $N_{si}(\tau)$ as before and $P_{si}^{(2)}(0,\tau,0)$ with $N_{si}^{(2)}(\tau)$, the count rate for triple coincidences of an idler photodetection event at $t_i =0$, and two signal photodetection events at $t_1 \in [-\tau_{\rm coin}, \tau_{\rm coin}]$  and $t_2 \in [\tau-\tau_{\rm coin}, \tau+\tau_{\rm coin}]$. In our HBT interferometer, we can equivalently look at the number of triple coincidences on the idler and $s_1$-$s_2$ photodetectors. By accounting for the resolution of the three photodetectors involved in our measurement, we obtain \cite{note1}
\begin{equation}
N_{si}^{(2)}(\tau)  = \frac{1}{(2 \tau_{\rm coin})^2 }\int_{-\tau_{\rm coin}}^{\tau_{\rm coin}}{d t_1 \int_{\tau-\tau_{\rm coin}}^{\tau+\tau_{\rm coin}} {d t_2 \bar P_{si}^{(2)}(t_1,t_2,0)}},
\end{equation}
where
\begin{eqnarray}
\bar P_{si}^{(2)}(t_1,t_2,0) &=& \int{d t_i \int{d t_{s_1} \int{d t_{s_2} u(t_i) u(t_{s_1}-t_1)}}} \nonumber\\ &\,&  \times \, \, u(t_{s_2}-t_2) P_{si}^{(2)}(t_{s_1},t_{s_2},t_i)
\end{eqnarray}
is the multi-coincidence rate for detecting an idler photon at time $0$ and two signal photons at times $t_1$ and $t_2$.

\begin{figure}
\centering
\includegraphics [width=.85\linewidth] {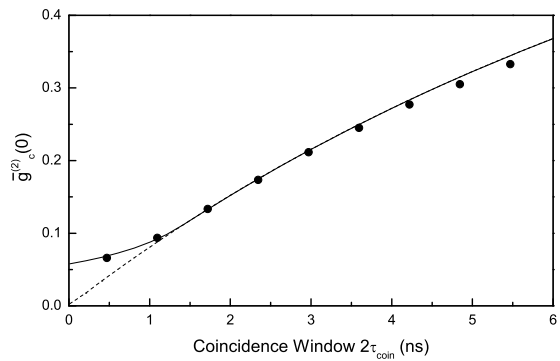}
\caption{{{(Color online) Experimental (symbols) and theoretical (line) results for the minimum of the time-averaged conditional coherence function, ${\bar g_{c}^{(2)}(0)}$, as a function of the coincidence window $2\tau_{\rm coin}$ using the same set of parameters as in Fig.~\ref{Fig:g2corr}. The dashed line is for ideal photodetectors ($\tau_d = 0$).
}}\label{Fig:window}}
\end{figure}

Figure \ref{Fig:g2dip} shows our measurement results for the time-averaged conditional coherence function $\bar g_c^{(2)}(\tau) \equiv {N_{si}^{(2)}(\tau) R(0)}/{ [N_{si}(0) N_{si}(\tau)]}$ for three different coincidence windows, which result in three different widths for the observed central dip. Here, $R(0)$ is obtained from the idler count rate in the experiment. As explained in Fig.~\ref{Fig:g2corr}, the ringing structure in Fig.~\ref{Fig:g2dip} is caused by double optical reflections. The graphs, nevertheless, exhibit the signature of a good SPS as the measured value of $\bar g_c^{(2)}(0)$, at 43$\,$MHz single-photon generation rate, in Fig.~\ref{Fig:g2dip}, is $0.0781 \pm 0.0006$ for $2 \tau_{\rm coin} = 0.78 \,$ns and $\tau_d = 0.35\,$ns. By reducing the pump power we can reduce this residual $\bar g_c^{(2)}(0)$ almost arbitrarily at the expense of reducing the total count rate. To see how the depth of the dip in Fig.~\ref{Fig:g2dip} varies with the coincidence window, in Fig.~\ref{Fig:window}, we have plotted $\bar g_c^{(2)}(0)$ versus $2\tau_{\rm coin}$. It can be seen that, for $\tau_{\rm coin} \ll \tau_d$, $\bar g_c^{(2)}(0)$ is determined by $\tau_d$, whereas, for $\tau_{\rm coin} \gg \tau_d$, it is almost linearly increasing with $\tau_{\rm coin}$. Our theoretical treatment is again well capable of reproducing the measurement results. The graph shown in Fig.~\ref{Fig:window} exemplifies the fact that a single value for $\bar g_c^{(2)}(0)$ does not bear enough information to quantify the source performance. At a fixed rate, the interplay between the coincidence window and the time resolution of photodetectors must also be accounted to give a proper figure of merit for an SPS. 

In this Report, we theoretically and experimentally studied the coherence properties of heralded single-photon sources that use parametric down-conversion. We used the Gaussian characteristics of down-converted fields to analytically find the degree of second-order coherence between signal and idler fields as well as for the signal field, individually, when it is conditioned on the detection of an idler photon. Our theory is well capable of reproducing our experimental results, which demonstrated a high-quality source of sub-picosecond single photons. It also allowed us to study the impacts of the chosen coincidence window, the down-conversion parameters, and the resolution of photodetectors on the outcome. Such analysis enables proper standardization of such devices. We would like to thank N. L\"utkenhaus, I. S\"ollner, and A. Safavi-Naeni for their technical assistance and acknowledge NSERC, CFI, ORF-RI, ORDCF, QuantumWorks, CIPI, and CIFAR for their financial support.

\begin{thebibliography}{99}
\bibitem{note1} Throughout the Letter, we assume unity quantum efficiencies for all employed photodetectors. Due to their normalized structures, coherence functions do not depend on photodetectors' quantum efficiencies.
\end{thebibliography}


\end{document}